\documentclass[a4paper,12pt]{article}

\usepackage{amssymb,amsmath}
\usepackage[pdftex]{graphicx}
\usepackage{color}
\usepackage[utf8]{inputenc}
\usepackage{indentfirst}
\newcommand{\ZZ}{\mathbb Z}

\newcommand{\ord}[1]{\mbox{\boldmath{$\mathcal{O}$}}\left({#1}\right)}

\def\limar#1#2{\,\raise0.3ex\hbox{$\longrightarrow$\kern-1.5em\raise-1.1ex
\hbox{$\scriptstyle{#1\rightarrow #2}$}}\,}

\def\limarr#1#2{\,\raise0.3ex\hbox{$\longrightarrow$\kern-1.5em\raise-1.3ex
\hbox{$\scriptstyle{#1\rightarrow #2}$}}\,}

\def\limlar#1#2{\ \raise0.3ex
\hbox{$-\hspace{-0.5em}-\hspace{-0.5em}-\hspace{-0.5em}
\longrightarrow$\kern-2.7em\raise-1.1ex
\hbox{$\scriptstyle{#1\rightarrow #2}$}}\ \ }

\title{A Quality and Cost Approach for Comparison of Small-World Networks}
\author{A.~Demichev$^{1,2}$, V.Ilyin$^{1,2}$, A. Kryukov$^{1,2}$ and S.Polyakov$^2$\\
{\small\textit{$^1$National Research Centre "Kurchatov Institute"}}\\ 
{\small\textit{$^2$Skobeltsyn Institute of Nuclear Physics, Lomonosov Moscow State University}}}
\date{}

\begin{document}

\maketitle
\begin{abstract}
We propose an approach based on analysis of cost-quality tradeoffs for comparison of efficiency of various algorithms for small-world network construction.  A number of both known in the literature and original algorithms for  complex small-world networks construction are shortly reviewed and compared. The networks constructed on the basis of these algorithms have basic structure of 1D regular lattice with additional shortcuts providing the small-world properties. It is shown that networks proposed in this work have the best cost-quality ratio in the considered class.

\end{abstract}


\section{Introduction\label{Vved}}

There exists a number of important cases of real world networks which can be modeled by a regular lattice with additional long-range connections between nodes (shortcuts).  While the local connectivity structure for such a networks remains similar to the original underlying lattice, the shortest path between any two distant lattice nodes becomes much smaller so that the resulting network possesses the small-world property with at most logarithmic increase of the mean distance between nodes with growth of the system size (see, e.g. \cite{Bar} and refs therein). Examples of such networks includes linear polymer networks, transportation networks, \textit{in vitro} neuronal networks (neuronal cultures), lattice models for space-time with wormholes, mobile and wireless networks and interconnection networks for supercomputers. For definiteness, in this work we have in mind as a principal use case designing an optimal interconnection networks for future generation supercomputers though the results obtained can be applied to other real networks of the type considered.

After the Petaflop ($10^{15}$~FLOPS) performance barrier for supercomputers was broken, the high-performance computing  community is exploring development of the principles for creation of next generation systems with performance of the order of Exaflop ($10^{18}$~FLOPS) \cite{SSG}. Though first Exaflop systems are estimated to be built between 2018 and 2020, approaches and basic principles of their construction are currently being actively investigated because on a way to their creation it is necessary to overcome a number of complex scientific and technical challenges and to elaborate essentially new approaches to development of their architecture and the hardware implementation.

One of the most important components of any supercomputer is the interconnection network which primarily defines ability to increase the number of computing nodes that is necessary for achieving the desirable performance. According to very modest estimations (see, for example, \cite{INW}), interconnection networks of Exaflop supercomputers must have more than 100 000 endpoints/CPUs (each most likely with many cores). Thus, one of key problems which should be solved on a way to next generation supercomputers is the development of interconnection network architecture with good scalability and ability to provide communication between huge number of computing nodes.

Three basic aspects of any interconnection network which mostly define its functionality are:
\begin{itemize}
\item network topology,
\item flow control, and
\item routing algorithm.
\end{itemize}

In this work we discuss mainly network topology. Two other aspects are quite important but out of the scope of this paper. A correct choice of network topology is vital for network design because both flow control and routing algorithms are heavily based on the topological properties of the network under consideration. Moreover, we shall consider only so called direct interconnection networks in which every node serves both as a processor unit and as a router. Indirect networks have their own merits for relatively small networks but are not well scalable. Among direct networks the interconnection networks with lattice structure and $D$-dimensional torus topology are widely used. In supercomputing literature such networks are termed ``$k$-ary $n$-cube'' \cite{DT} ($n$ is the torus dimension, in our notations $n = D$; $k$ is the number of nodes along each dimension). These networks possess a number of advantages, in particular:
\begin{itemize}
\item high resilience of the networks as a whole in case of failure of some nodes (existence of bypass communication paths), 
\item hypercube of dimensionality which is enough for a given number of nodes provides small average distance between nodes, 
\item in case of simulation of $D$-dimensional objects, the structure of the computing tasks is mapped to the $D$-dimensional lattice of computing nodes in an optimal way; it is worth mentioning that just tasks of this kind (especially for $D=3$) are supposed to constitute a considerable part of tasks for the next generation supercomputers.
\end{itemize}

However in the case of huge number of nodes typical for next generation supercomputers, the $D$-dimensional lattice toruses have also essential shortcomings. In particular, lattices of low dimensionality have very large average path length between nodes while lattices of high dimensionality, comparable with logarithm of number of nodes, are inconvenient for embedding into real physical space and therefore for implementation because of huge length of communication links. On the other hand, it is well known that interconnection networks with minimal average distance for a given number of nodes have the best operation characteristics, such as, e.g., performance and robustness (see, for example, \cite {Kle}). Thus usual networks with simple structure of the regular lattices will be ineffective for solution of tasks of general type except those related to triangulation of $D$-dimensional objects.

For that matter it seems attractive to use for the next generation supercomputers interconnection networks with small-world property. The point is that while for the regular $D$-dimensional lattice the average distance $d$ between nodes grows as the power of number of nodes: $d\sim N^{1/D}$, in the case of a network with the small-world properties $d$ grows essentially slower: $d\sim \ln\,N$. It is worth mentioning that some complex networks have average distance that depends on the number of nodes in the power-like manner similarly to the regular lattices but with the exponent which is significantly smaller: $d\sim N^\gamma,\  \gamma\ll 1/D$. Quite often investigation of network properties are carried out by means of numerical simulations and it is not easy to distinguish such power-like behavior from logarithmic one. From the applications point of view and in the case of small enough exponent the difference is unessential too. Therefore we will consider that such networks also possess the small-world property.

\begin{figure}
\begin{center}
\includegraphics[scale=.3]{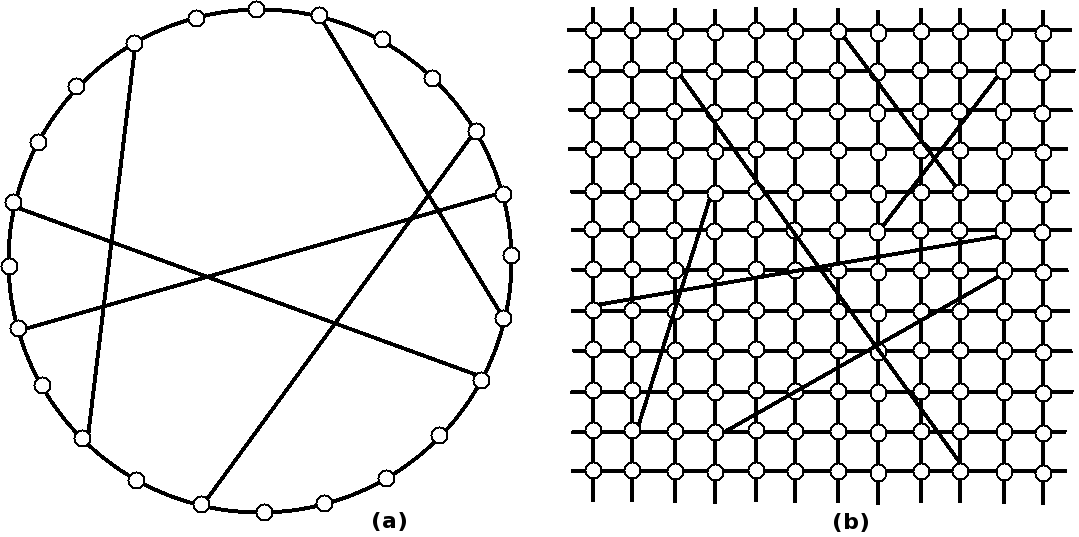}
\end{center}
\caption{A simplified form of the lattice networks supplemented with shortcuts in one-dimensional~(a) and two-dimensional~(b) cases}
\label{fig:RSP_1D_2D}
\end{figure}

In the classical version \cite{WS} complex networks with small-world properties appears at the intermediate stage in the process of stochastic transformation of a regular lattice into Erd\"os-R\'enyi random graphs \cite {ER}, \cite{AB}, the structure of the regular lattice is being broken. As it was mentioned above, this is undesirable for the interconnection networks. Therefore in this work we shall use the modification of the small-world networks construction in which the original lattice is preserved and is supplemented by additional links (shortcuts) providing the small-world properties (see \cite{Bar}, sect. IV.C.2 and refs. therein). A simplified form of such networks in one- and two-dimensional cases is depicted in fig.~\ref{fig:RSP_1D_2D}. Below for brevity we shall use for such networks the term ``lattice networks with shortcuts'' (LNS).

It is necessary to note that besides the small average distance between nodes one more distinctive property of the small-world networks is high degree of clusterization \cite{WS}, \cite{AB}. High clusterization provides local fault tolerance of the networks because of existence of local bypass paths in the case of a node failure. Note, however, that in the case of LNS with $D > 1$ such a local fault tolerance (existence of local bypass paths) is provided by the underlying lattice.

Path length (distance) between nodes is understood in the network sense: as the minimum number of edges which it is necessary to pass to get from one node to another. Respectively the average distance between nodes is defined as an average over all pairs of nodes for a given network. However for huge interconnection networks such an average distance may prove to be inadequate characteristic. The point is that for finding the shortest routes it is necessary to know global structure of the network (positions of all the shortcuts). Therefore the routing of messages using the shortest paths can be too cumbersome and ineffective because it requires storing and processing large data amounts. Thus routing algorithms based on local navigation \cite{Mil}, \cite{K2000} become of special importance. 

In case of local navigation the node ``knows'' only a geographical position (position in the underlying lattice) of all nodes and the nearest neighbors with the account of shortcuts. Using only this information (information about all the shortcuts is not used) it is necessary to deliver a message to the destination node along shortest possible path. The simplest solution of this problem is to use the greedy algorithm: the current node sends the message to that of its neighbors which geographically (in the sense of coordinates on the underlying lattice) is closest to the destination node. Later we will consider a more complicated modification for the local navigation algorithm.

In this work we investigate the average global and local navigation paths between nodes as the most important specifications defining communication properties of networks. The main goal is to develop an optimal algorithm for construction of a network with large number of nodes but small average global and/or local navigation path lengths between nodes. A general idea is to add to a regular lattice shortcuts according to an algorithm (or algorithms) in such a way that ``cost-quality'' ratio be optimized for the network obtained. Under the character of ``cost'' we choose unit wiring cost, i.e., total length of all shortcuts divided by the number of nodes in the network (total shortcuts length per node), while ``quality'' is the global or navigation path length between nodes. 

As it was already mentioned, the original algorithm \cite{WS} for small-world network construction is stochastic one: at each step of the algorithm the graph edges randomly change their positions with a given probability. As result of such an algorithm there appears an ensemble of graphs with some distribution of their characteristic properties, in particular, with some distribution of average global and local navigation path lengths between nodes. For many real networks the stochasticity is an inherent property (e.g., this is true for Web and Internet; other examples see, e.g., in \cite{AB}). However in the case of interconnection networks (as well as for other transportation networks, e.g. for airline networks) the design/construction process is under full control of a developer. Therefore the stochasticity is not inherent for this process. Thus an important question is the following: is there such a deterministic algorithm for lattice modification turning it into a small-world network that the cost-quality ratio be better than in the case of stochastic algorithms. Actually obtaining an answer to this question is one of the principal goals of this work.

Though some results are valid for lattices of any dimension, in this work we consider the simplest case of one-dimensional lattices with torus topology (see fig.~\ref{fig:RSP_1D_2D}(a)). The main reason for this is a convenience of optimization of various algorithms and of comparison of networks properties because of shorter simulation time. Generalization to high-dimensional lattices will be considered in subsequent papers.

This paper is organized as follows. In the next section \ref{MMR} a short review of both known in the literature and original algorithms of LNS construction is presented. In section \ref{SDMU} general properties of global average distance of the obtained LSP are considered while section \ref{LNRP} is devoted to consideration of average navigation path length. In section \ref{OAKS} we present a method for selection of best lattice network with shortcuts (LNS) on the basis of cost-quality ratio optimization. The section \ref{Zak} contains a conclusion.

\section{Small world on a lattice: construction and main properties\label{MMR}}

\subsection{Algorithms for shortcuts establishing\label{APPO}}

\subsubsection{Stochastic algorithms}
The stochastic algorithm of small-world network construction preserving the underlying lattice was suggested in a number of works, in particular, \cite{MM}, \cite{SC}, \cite{PR2006} (see also the review \cite{Bar} and refs. therein). In one-dimensional case it is formulated below.

\paragraph{Algorithm S1\label{S1}}
\begin{enumerate}
\item \label{ALG1step1} The starting network is a one-dimensional lattice with $L$ nodes and topology of a circle;
\item\label{ALG1step2} 
one goes sequentially through all the network nodes and attach to each node $i$ the first end of a shortcut with probability $0<p\leq1$;
\item \label{ALG1step3}
the second shortcut end (that is the node $j$ to which this end is attached) cannot coincide with the neighbors of the first end in the sense of the underlying lattice and cannot result in duplication of an existent shortcut; in all other respects it is chosen randomly with probability
\begin{equation}
P(r)\sim r^{-\alpha}\ ,                                    \label{raspPerem}
\end{equation}
which is the power-like function of the lattice distance $r=r_{ij}$ between the nodes $i$ and $j$. 
\end{enumerate}

An ensemble of LNSs constructed by means of the algorithm S1 is parametrized by $L,\ p$ and $\alpha$.  With the fixed network size $L$, the parameters $p$ and $\alpha$ should be optimized to achieve the best cost-quality ratio. A number of the shortcuts in the ensemble is distributed according to the binomial distribution (a number of successful ``tests'' in the $L$ trials) so that the average number of the shortcuts is $pL$. 

In the search for the possible interconnection network structure the following modification of the algorithm S1 may be useful.

\paragraph{Algorithm S1m\label{S1m}}
\begin{enumerate}
\item \label{ALG2step1} The starting network is a one-dimensional lattice with $L$ nodes and topology of a circle (as in the case of S1);
\item\label{ALG2step2} 
a number of shortcuts $t$ which have to be added to the lattice is fixed;
\item\label{ALG2step3} 
out of all the $L$ lattice nodes one randomly chooses $t$ nodes and attaches to them first ends of shortcuts;
\item \label{ALG2step4}
the second end of each shortcut is chosen randomly and identical to the step \ref{ALG1step3} of the algorithm S1.
\end{enumerate}

An ensemble of LNSs constructed by means of the algorithm S1m is parametrized by $L,\ t$ and $\alpha$. The distinction from the previous basic algorithm is that in the case of S1 both positions and the number of shortcuts are random entities while in the case of S1m only positions of the shortcuts are chosen randomly. In other words, the degree of stochasticity of the algorithm S1m becomes lower because of the contraction of the probabilistic sample space (the set of actual LNS instances). On the other hand, sum of the sample spaces for all allowable values of $t$ in the case of S1m is equivalent to the sum of the sample spaces for all allowable values of $p$ in the case of S1. Therefore the search for an optimal LNS created according to the algorithms S1 or S1m have to lead to the same results under condition of large enough representative sampling.

Since our principal use case concerns optimal large interconnection network for supercomputers (which intrinsically is not stochastic) in our case the algorithm S1m is more preferable than S1.

The next algorithm is the more essential modification of the algorithms S1 and S1m and further reduces the degree of stochasticity.

\paragraph{Algorithm S2\label{S2}}
\begin{enumerate}
\item \label{ALG3step1} The starting network is a one-dimensional lattice with $L$ nodes and topolgy of a circle (as in the case of S1 and S1m);
\item\label{ALG3step2} 
a number of shortcuts $t$ which have to be added to the lattice is fixed (similarly to S1m) and a fraction $c < t$ of the shortcuts which should be added in special way is fixed;
\item\label{ALG3step3} 
out of all the $L$ lattice nodes one randomly chooses $(t-c)$ nodes and attaches to them first ends of shortcuts;
\item \label{ALG3step4}
the second end of each of the $(t-c)$ shortcuts is chosen randomly and identical to the step \ref{ALG1step3} of the algorithm S1;
\item \label{ALG3step5} 
$c$ shortcuts are added one by one and at each step:
\begin{itemize}
\item \label{ALG3step6} 
one updates the list of all nodes to which shortcuts have been already added (taking into account both the shortcuts added at the previous steps \ref{ALG3step2}–\ref{ALG3step4} and at the current step of the algorithm);
\item 
from the list obtained one node is randomly chosen and a new shortcut is added to it;
\item 
the second end of each of this additional $c$ shortcuts is chosen randomly and identical to the step \ref{ALG3step4} (any new shortcut should not coincide with any of previously established or with an underlying lattice edge).
\end{itemize}
\end{enumerate}

An ensemble of LNSs constructed by means of the algorithm S2 is parametrized by $L,\ t,\ c,\ \alpha$. In the algorithm S2 not only the number of shortcuts is fixed but also their positions are not quite random but rather subjected to an additional rule the degree of this regularity being determined by the parameter  $c$ (at $c = 0$ the algorithm S2 is equivalent to S1m). In a sense S2 is a hybrid algorithm, i.e., partially stochastic and partially deterministic.

The simulation shows that the average distance between nodes in case of S2 is smaller than for S1/S1m for wide range of the parameter $c$.

Notice that there possible other modifications of the stochastic algorithms. For example, in \cite{LRMHSA}, \cite{LZHZD} an algorithm with fixed total length of shortcuts is suggested. Such a version may be adequate to the case of working out of transport networks under condition that only limited resources (e.g., financial) are available for real implementation of the shortcuts. However, in the use cases similar to the interconnection network designing a more adequate approach, in our opinion, is the multicriteria  optimization of the cost-quality ratio which will be discussed in section \ref{MOCK}.

\subsubsection{Deterministic algorithms\label{DetAlg}}

A most natural attempt of LNS construction with the small-world properties by means of a deterministic algorithm is based on the simple idea: at each step of the algorithm a shortcut joins two most distant (in the sense of the existent by this step network) nodes.

\paragraph{Algorithm D1\label{D1}}
\begin{enumerate}
\item \label{ALG_D1_step1} The starting network is a one-dimensional lattice with $L$ nodes and topology of a circle;
\item\label{ALG_D1_step2} 
a number of shortcuts $t$ which have to be added to the lattice is fixed;
\item\label{ALG_D1_step3} 
one goes sequentially through all the pairs of the network nodes and chooses the ones separated by maximal network distance (i.e., with the account of shortcuts already added);
\item \label{ALG_D1_step4}
from the set selected at the preceding step one chooses a pair with shortest lattice distance and joins it by a shortcut;
\item \label{ALG_D1_step5} the steps 3 and 4 are repeated $t$ times.
\end{enumerate}
A version of this algorithm was discussed in \cite{ZMHGY}. Besides the lattice size $L$, this algorithm contains the only parameter $t$, which can be used for the cost-quality optimization.

\vspace{3 mm}

The next deterministic algorithm was suggested in \cite{BGA}, \cite{BGG} and named there by ``hierarchical HN4 algorithm''. 

\paragraph{Algorithm D2 (HN4)\label{D2}}$ $
 
\begin{enumerate}
\item \label{ALG_D2_step1} The starting network is a one-dimensional lattice with $L=2^k$ nodes and topology of a circle;
\item\label{ALG_D2_step2} 
the nodes are labeled by integers $n\ (0 < n \leq L)$ which, in turn, are parametrized by two integers $\{i,j\},\ 0\leq i\leq k$ as follows:
\begin{equation}
n=2^i(2j+1)\  ;                                               \label{MMR01}
\end{equation}
\item\label{ALG_D2_step3} 
for any given $i$ ($0\leq i\leq k$) defining a level of the hierarchy the closest (in the sense of the underlying lattice) neighbors are connected by shortcuts (i.e., a node with the number $2^i(2j+1)\{\mbox{mod}\ L\}$ is connected with the nodes $2^i[2(j-1)+1]\{\mbox{mod}\ L\}$ and $2^i[2(j+1)+1]\{\mbox{mod}\ L\}$).
\end{enumerate}
Following the general idea of the D2-algorithm one can construct LNSs in which the the hierarchy levels are defined by powers of other (not of 2) integers.

The LNSs, constructed by the D2-algorithm, are reminiscent of the well-known circulant graphs (see, e.g., the review \cite{Mon} and refs. therein). Some subclass of the circulant graphs, namely the multiplicative circulant graphs, also have the small-world properties. Such graphs are denoted as follows: $C(s^k;1,s,s^2,\dots,s^{k-1})$, where $s^k=L$ is the network size, the unity corresponds to existence of the underlying lattice and $s,s^2,\dots,s^{k-1}$ are lengths of the shortcuts. In particular, for diameter of these graphs one has (\cite{Mon} and refs. therein):
\begin{eqnarray}
{\cal D}&=&k\lfloor s/2\rfloor \qquad \mbox{for odd}\qquad s>1\,,\nonumber\\
\label{circDiam}\\
{\cal D}&=&ks/2-\lfloor k/2\rfloor \qquad \mbox{for even}\qquad s>1\,.\nonumber
\end{eqnarray}
Here $\lfloor x \rfloor$ is the largest integer not greater than $x$.

However the circulant networks have large total wiring cost per node which is equal $(s^k-1)/(s-1)$. The next deterministic algorithm, suggested in  \cite{COP} and \cite{CMP}, uses the circulant networks with short-range shortcuts as starting ones and then supplement them with a long-range shortcuts which provide the small-world properties. Since detailed formulation of this algorithm is rather lengthy we present here only its basic ideas. The details can be found in the above cited works.

\paragraph{Algorithm D3\label{D3}}
\begin{enumerate}
\item The starting network is a one-dimensional lattice with $L$ nodes and topology of a circle or the circulant graph $C(L;1,\dots,K/2)$ (the usual lattice is equivalent to the circulant graph with $K=2$);
\item one chooses a subset of nodes from the starting network (the nodes from this subset are called hubs); 
\item the hubs are connected by edges (shortcuts) so that this results in some graph $H$, whose nodes are the hubs and edges are the added shortcuts;
\item in addition, one carries out a special local reorganization of the edges so that degrees of all the nodes be the same. 
\end{enumerate}

On the basis of the ideas of the D2-algorith and the general structure of the multiplicative circulant graphs, in this work we suggest a new algorithm providing logarithmic dependence of the average distance on the network size. Similarly to the multiplicative circulant graphs, the shortcuts of the lengths $b, b^2, \ldots, b^k$ ( for some $b,\ k\in \ZZ$) are added. However, the number of the shortcuts decreases with their length so that unit wiring cost grows proportionally to $k$ but not to $b^k$. At first, we formulate a simplified version of such an algorithm.

\paragraph{Algorithm D4s\label{D4s}}
\begin{enumerate}
\item The starting network is a one-dimensional lattice with $L=mb^k$ ($m\in \ZZ$ nodes and topology of a circle;
\item starting from an arbitrary node all the network nodes are labeled by numbers from $0$ to $L-1$;
\item the nodes $0, b, 2b, \ldots, L-b$ are connected by shortcuts in a cycle (the node $0$ is connected with $b$, $b$ with $2b$, etc.; $L-b$ is connected with $0$);
\item for each $i=2, \ldots, k$ the nodes $0,\ b^i,\ 2b^i,\ \ldots,\ L-b^i$ are connected in a cycle.
\end{enumerate}

\begin{figure}
\begin{center}
\includegraphics[scale=.5]{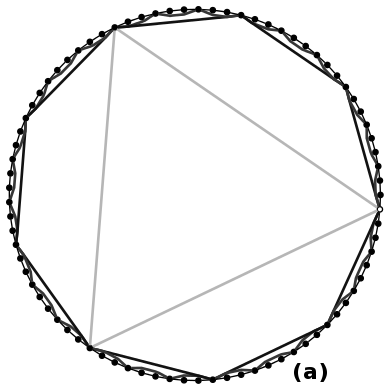}
\hspace{10mm}
\includegraphics[scale=.5]{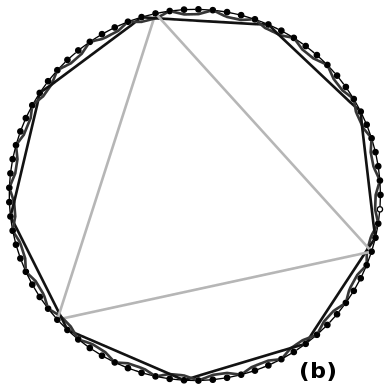}
\end{center}
\caption{LNSs constructed by the algorithm D4s (a) and by D4 (b); in both cases $L=81,\ b=3,\ k=3$; the node labeled by the number ``0'' is depicted by empty circle}
\label{fig:D4s-D4}
\end{figure}

The networks constructed by this algorithm has the following structure (see fig.~\ref{fig:D4s-D4}(a)): the end points of each of $m$ shortcuts of the length $b^k$ are connected by a chain of $b$ shortcuts of the length $b^{k-1}$, the endpoints of the latters are connected by a chain of $b$ shortcuts of the length $b^{k-2}$, etc. It is easy to show that even for the simplest greedy navigation algorithm the maximal path length between any two nodes and therefore the average distance have the order  $\ord{(kb+m)}$. This implies that for fixed  $b$ and $m = L/b^k = O(1)$ the average distance proved to be $\ord{\log L}$. More detailed estimations are presented in section  \ref{SDDA}.

We will call the networks constructed by this algorithm subcirculant ones. The shortcoming of the subcirculant networks constructed by the algorithm D4s is the low robustness with respect to possible faults at the nodes $0,\ b^k,\ \ldots,\ (m-1)b^k$: since there are no bypass paths around these nodes a fault of any of them may considerably increase load (betweenness centrality) for other nodes. Moreover removal of a couple of these nodes makes the network a disconnected one. Besides the nodes of the highest level (i.e., the nodes with longest shortcuts) have high betweenness centrality and in the case of large network size plus relatively small $b$ they have high node degree. These properties may cause problems in case of real implementations.

In the following modification of the algorithm D4s the mentioned problems are solved by parting the cycles of the different levels to the neighboring (in the sense of the underlying lattice) nodes. In addition, the modified algorithm is generalized to the case of arbitrary network size $L$ (not necessarily divisible by $b$).

\paragraph{Algorithm D4\label{D4}}
\begin{enumerate}
\item The starting network is a one-dimensional lattice with $L$ nodes and topology of a circle;
\item positive integers $b,\ k$ satisfying the conditions
\begin{eqnarray}
2 < b \leq L/2\ ,                              \label{D4-1}\\
b^k+k < L\  ,                                  \label{D4-2} \\
k \leq b^2\ ;                                  \label{D4-3}
\end{eqnarray}
are chosen; 
\item starting from an arbitrary node all the network nodes are labeled by numbers from $0$ to $L-1$;
\item $k-1$ cycles are created in the following way: for $i=2,\dots,\ k-1,\ k,$ the nodes $i,\ i+b^i,\ i+2b^i,\dots,\ i+h_i b^i$ where  $h_i=\lfloor (L-i)/b^i \rfloor$ are consecutively connected by shortcuts in a cycle; 
\item one more cycle is created:
\begin{itemize}
\item for $j=0, \ldots, h_1$ where $h_1=\lfloor (L-1)/b\rfloor$ one selects the nodes $1+jb$ and then for each node chooses the nearest in order of the labeling node $q_j$ which does not have so far a shortcut; in such way on obtains a set of nodes  ${\cal C}_1^\prime=\{q_0,\ q_1,\dots,\ q_{h_1}\}$;
\item if this set contains neighbors in the sense of the underlying lattice, i.e. $\exists\ \{q_j,\ q_{j+1}\}:\ q_{j+1}-q_j=1$ then the node $q_j$ is removed from the set; the resulting set is denoted ${\cal C}_1$;
\item the nodes from the set ${\cal C}_1$ are consecutively connected by shortcuts in a cycle.
\end{itemize}
\end{enumerate}

The D4 LNS structure provides that the local navigation algorithm comparing the lattice distance to a destination node for all neighbors of neighbors of a source node proves to be comparable in its effectiveness with the algorithm using globally shortest paths. In more details the properties of the D4 LNSs are discussed in sections \ref{SDMU} and \ref{LNRP}.

Notice that the condition (\ref{D4-3}) is not absolutely necessary and is used here for simplicity: if $k>b^2$ one has to varies lengths of some shortcuts in more than one cycle in order to consecutively put $k$ nodes entering $k$ different cycles.

A sample of the graphs constructed by the algorithm D4 for $L=81,\ b=3,\ k=3$ is depicted in fig.~\ref{fig:D4s-D4}(b). Of course, the samples in this figure are merely illustrative: actually we are interested in networks of essentially larger size. In particular, in the following sections we will present results for such networks with $L\geq 10^4$. 

\section{Average distance between nodes \label{SDMU}}

\subsection{Average distance for stochastic LNS\label{SDEA}}
With the use of the stochastic algorithms the average path length $d$ between nodes is a result of the double averaging:
\begin{itemize}
\item over the statistical ensemble of random graphs, obtained as a result of the the stochastic process of shortcuts establishing; this averaging is denoted by 	angular brackets: $\langle\cdot\rangle$;
\item over all the pairs of nodes for each instance of the network:
\end{itemize}
\begin{equation}
d= \frac{2}{L(L-1)}\sum_{i>j}\left\langle {d}_{ij}\right\rangle=\frac{1}{L(L-1)}\sum_{i,j}\left\langle {d}_{ij}\right\rangle\ .   \label{SDEA03} 
\end{equation}

\begin{figure}
\begin{center}
\includegraphics[scale=.4]{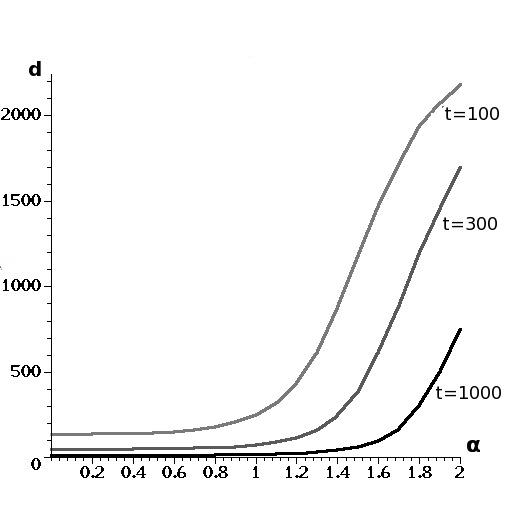}
\end{center}
\caption{The dependence of the average path length on LNS's parameters}
\label{fig:srDlina_ot_alpha_mod2}
\end{figure}

The typical dependence of the average distance between nodes for LNSs, created by the algorithm S1m, on the parameters $\alpha$ and $t$ is presented in fig.~\ref{fig:srDlina_ot_alpha_mod2}. The results are obtained by numerical simulation for an ensemble of 1000 instances with $L = 10^4$ nodes. For $\alpha > 2$ the average shortcut length becomes too small and they do not influence the network properties so that the small-world properties disappear \cite{MM}, \cite{PR2006}. 

\begin{figure}
\begin{center}
\includegraphics[scale=.6]{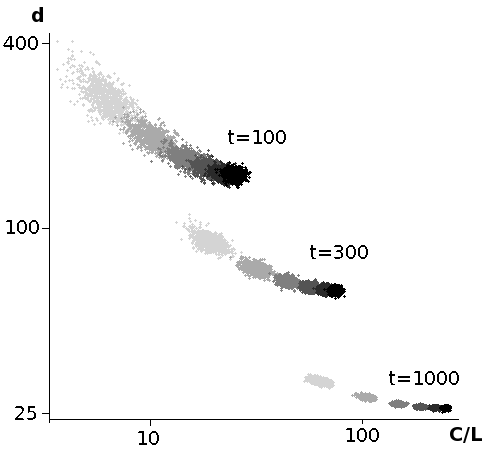}
\end{center}
\caption{Dependence of the distances between nodes averaged over instances of the network on the total shortcuts length per node}
\label{fig:d_of_C_L10000_t100to1000_alpha0to1}
\end{figure}

From naive consideration of the results depicted in fig.~\ref{fig:srDlina_ot_alpha_mod2} (for the algorithm S1 and the parameter $p$ the dependence is quite similar) one may conclude that to construct a network with the best value of the average distance between nodes it is enough to choose the largest possible value of $t$ (or $p$) and the smallest value of $\alpha$. However, as it was stressed in the Introduction, each solution has its cost and one has to optimize the cost-quality relation for sought network (otherwise the obvious solution is just the complete graph in which all the nodes are connected by direct links). Since in this work the total shortcuts length per node serves as the ``cost'' one has to analyze dependence of the mean distance on this quantity. Typical results of the numerical simulation for such a dependence are depicted in fig.~\ref{fig:d_of_C_L10000_t100to1000_alpha0to1}. 

In fig.~\ref{fig:d_of_C_L10000_t100to1000_alpha0to1} every point with the respective value in gray scale corresponds to an  instance of LNS in ensembles with $L=10^4$ nodes, $\alpha=0, 0.2,\ 0.4,\ 0.6,\ 0.8,\ 1.0$ (the more the value of $\alpha$ the lighter the shading) and with $t = 100,\ 300,\ 1000$; the vertical axis corresponds to the average (over instance) distance between nodes while the values on the horizontal axis shows actual (numerically calculated) total shortcuts length per node $C/L$ ($C$ is the total shortcuts length, $L$ is the number of nodes). The logarithmic scale is used for both axis. In the case of the algorithm S1, a similar results were obtained but with wider distribution (as expected). These results imply that for decreasing the average distance at equal cost one have to choose an ensemble with maximally possible value of $t$ (or $p$) and as well with maximally possible for the given cost value of $\alpha$. In other words, large number of not too lengthy shortcuts is a cheaper solution.

The results depicted in fig.\ref{fig:d_of_C_L10000_t100to1000_alpha0to1} and as well similar results obtained for other values of the parameters give a more detailed representation of the dependence of the average distance on the cost than the results of the work \cite{PR2006}. In the latter the results for distance averaged both over instance and over ensemble is presented. Besides the ``theoretical'' total shortcut length $C_W/L = p\bar{s}$ where $\bar{s}$ is the shortcut length averaged with respect to the distribution (\ref{raspPerem} is used as the wiring cost. In our work we use real total shortcut length numerically calculated for a given instance of the networks. The presented detailed data shows that direct using cost-quality relation does not provide unequivocal choosing an optimal network (i.e., in the case of algorithms S1 and S1m does not provide choosing the optimal values for $p$, $t$ and $\alpha$). Therefore one has to use methods of multicriteria analysis, see section  \ref{OAKS}.

\begin{figure}
\begin{center}
\includegraphics[scale=.4]{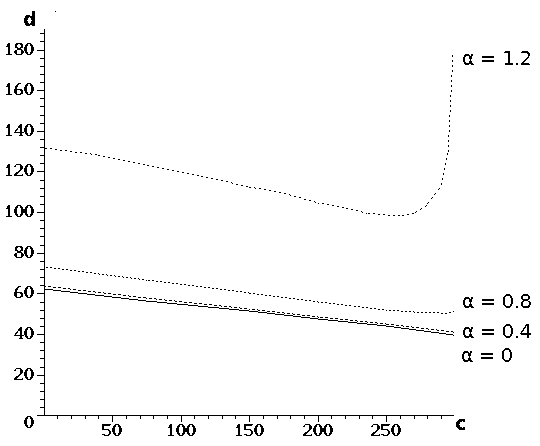}
\end{center}
\caption{Dependence of the average (over ensemble and instances) distance on the parameter $c$ (the number of forcibly conjoined shortcuts) in the case of the algorithm S2 for  $t=300,\ \alpha=0,\ 0.4,\ 0.8,\ 1.2$ and $L=10^4$}
\label{fig:d_of_c-S2}
\end{figure}

Numerical simulations show that average distances in LNSs, created by the algorithm S2, are smaller than in the case of S1/S1m in wide range of the parameter $c$ (the number of forcibly conjoined shortcuts), see fig.~\ref{fig:d_of_c-S2} .

Thus from the point of view of the global average path length, S2 proves to be preferred among all the stochastic algorithms. For S2 one can also find dependence of the average distance on total shortcut length per node (unit wiring cost) similarly to fig.~\ref{fig:d_of_C_L10000_t100to1000_alpha0to1}. However because of bigger number of parameters such a graphic presentation appears to be not quite instructive. Generally speaking, at the accounting of the wiring cost the algorithm S2 might be non-optimal even in the class of stochastic algorithms. However in the section \ref{MOCK} we shall show with the help of the appropriate method that S2 indeed most preferable stochastic algorithm.

\subsection{Average distance for deterministic LNSs\label{SDDA}}

The deterministic algorithms produce a unique instance of the network. Therefore the average distance between nodes is defined similarly to (\ref{SDEA03}) but without averaging over ensembles:
\begin{equation}
d=\frac{1}{L(L-1)}\sum_{i,j} {d}_{ij}\ .                                     \label{SDDA01} 
\end{equation}

\begin{figure}
\begin{center}
\includegraphics[scale=.5]{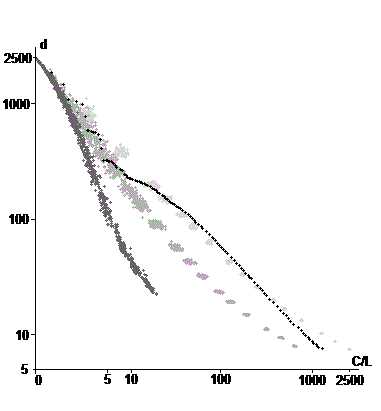}
\end{center}
\caption{Dependence of the average distance $d$ on the shortcuts length per node $C/L$ for one-dimensional networks with $L=10^4$ in case of the deterministic algorithm D1 (black dots) and for the stochastic algorithm S1m ($\alpha=0,\ 1,\ 2$, the darker the the dots the larger the values of $\alpha$); the number of shortcuts in D1 does not exceed $L$}
\label{fig:d_of_C_d1L10000_D1_vs_S1m_alpha0_1_2}
\end{figure}

\paragraph{LNS constructed by the algorithm D1.}
The behavior of the average distance between nodes of LNS, constructed by the algorithm D1, was investigated in \cite{ZMHGY} by means of numerical simulations. It was shown that with grows of the number of shortcuts $t$ the average distance decreases as a power low. Notice that a simple consideration of the quality-cost dependence allows in some cases to make a choice among different types of the algorithms. In particular, the dependence of the average distance on the shortcuts length per node for deterministic algorithm D1 and for stochastic one S1m is presented in fig.~\ref{fig:d_of_C_d1L10000_D1_vs_S1m_alpha0_1_2}.

The results in fig.~\ref{fig:d_of_C_d1L10000_D1_vs_S1m_alpha0_1_2} shows that the cost-quality ratio for the algorithm D1 is worse than that for S1m for large enough $\alpha$.
Thus S1m (or S1) is more preferable than D1. As it is clear from the discussion in the preceding section, this even more so for S2.

\paragraph{LNS constructed by the algorithm D2.} An estimation for the diameter of LNSs constructed by the algorithm D2 was obtained in  \cite{BGA}, \cite{BGG}: 
\begin{equation}
{\cal D}\sim (\log_2 L)^\beta\ , \qquad \beta\sim\frac{\sqrt{\log_2L^2}}{\log_2\log_2L^2}+\frac{1}{2}\ . 
\end{equation}

Strictly speaking, the exponent $\beta\limar{L}{\infty}\infty$ so that the diameter grows faster than any power of $\log_2L$. However $\beta$ grows very slowly ranging merely from $\beta\approx1.44$ to $\beta\approx1.84$ over nine orders of magnitude of varying $L$: from $L=10$ to $L=10^{10}$ \cite{BGG}. Therefore in practice it is possible to consider that such network possesses property of the small world. This is confirmed by results of our numerical simulations presented in the Table \ref{tab:D2-D4char}.

\begin{table}
\caption{Characteristic properties of LNSs constructed by D2 and D4}
\begin{center}
\begin{tabular}{|c|c|c|c|c|} \hline
$L$    & $C/L$ for D2 & $C/L$ for D4 & $d$ for D2 & $\ell^{(2)}$  for D4\\ \hline
1024   & 8.5          & 4.0          & 11.4563    & 12.4992 \\ \hline
4096   & 10.5         & 5.0          & 17.7199    & 17.0229 \\ \hline
16384  & 12.5         & 5.0          & 26.3773    & 21.8356 \\ \hline
65536  & 14.5         & 6.0          & 38.0082    & 27.0769 \\ \hline
262144 & 16.5         & 7.0          & 53.3045    & 32.4278 \\ \hline
524288 & 17.5         & 8.0          & 62.6001    & 35.6748 \\ \hline
\end{tabular}
\end{center}
\label{tab:D2-D4char}
\end{table}

The shortcuts length can be easily calculated: indeed, at each level $1,2,\ldots,k-2$ the shortcuts goes consequently one-by-one and encircle all the $L=2^k$ nodes of the underlying lattice. At the level $k-1$ there is one shortcut of the length $L/2$. As a result the total shortcuts length appears to be $L(k-3/2)$, and the length per node is $C/L = k-3/2$. 

However, the results presented in the table \ref{tab:D2-D4char} show that the subcirculant D4-networks have lower cost and for $L>4000$ have lower navigation length $\ell^{(2)}$ than the global average distance $d$ in case of the algorithm D2 (the superscript ``$(2)$'' in $\ell^{(2)}$ means that the two-level navigation algorithm is used, see section \ref{LNRP}). Since the global distance does not exceed any navigation path length, D4 is \textit{a fortiori} more preferred algorithm in comparison with D2.

\paragraph{LNS constructed by the algorithm D3.} Estimations for the diameter of networks constructed by the algorithm D3 were obtained in \cite{COP}. In particular,
\begin{equation}
{\cal D}\leq 2\lceil\left(\lceil L/h\rceil-1\right)/K \rceil +D_H \ .                   \label{D3-1}
\end{equation} 
Here $D_G$ is the diameter of the network with the shortcuts, $D_H$ is the diameter of the graph $H$, $h$ is the number of hubs; $\lceil x \rceil$  is the smallest integer not less than $x$. As the graph $H$ connecting the hubs, two types of graphs were considered: star graphs $S_{1,h-1}$ (one node is connected to all $h-1$ other nodes) with the diameter 2 and double loop circulant graphs $C(h;a,b)$ with the diameter
$$
D_{C(h;a,b)}= \lceil (-1+\sqrt{2h-1})/2\rceil \ .
$$

\paragraph{LNS constructed by the algorithm D4s.} Average distance between nodes in LNSs for the algorithm D4s does not exceed an average path length constructed by the following algorithm: 
\begin{itemize}
\item from the initial node $i$ move to the nearest node with the label $v_1 b$ through which passes a cycle made of shortcuts of length $b$ (let us call it the first level cycle); then using the first-level shortcuts move to the nearest node $v_2 b^2$ entering the second-level cycle and continuing in this way reach the node $N^k(i) = v_k b^k$ entering the top-level cycle;
\item a similar (though backward) path is created from the destination node $j$ to the node $N^k(j) = v_k' b^k$;
\item move from $N^k(i)$ to $N^k(j)$ by shortcuts of the length $b^k$ and then by the path constructed at the preceding step move to the destination node $j$.
\end{itemize}

Total length of all such paths is equal 
$$
2(L-1) \sum_{i=0}^{L-1} l_a(i) + \sum_{0\leq i, j < L} l_t(i,j) \ , 
$$
where $l_a(i)$ is length of the ``ascending'' path from $i$ to $N^k(i)$ and $l_t(i,j)$ is the number of the top-level shortcuts between $N^k(i)$ and $N^k(j)$.
For reasons of symmetry one finds that the number of nodes for which $N^k(i)=u b^k$ is equal to $b^k$ for each $u=0,1,\ldots,m-1$. Therefore
$$
\sum_{0\leq i, j < L} l_t(i,j) = m b^{2k} \sum_{u=0}^{m-1} l_t(0, ub^k) = 
m b^{2k} \left( \sum_{u=0}^{\lceil m/2 \rceil} u + 
                             \sum_{u=1}^{\lfloor m/2 \rfloor} u \right) \leq \frac{m^3 b^{2k}}{4} \ .
$$
The total length of ``ascending'' paths reads
$$
 k \left( \sum_{u=0}^{\lceil b/2 \rceil} u + 
                             \sum_{u=1}^{\lfloor b/2 \rfloor} u \right) \frac{L}{b} \ ,
$$
from which on finds that the total length of entire paths does not exceed
$(L-1)L k \frac{b}{2} + \frac{m}{4} L^2$ for even $b$ and  
$(L-1)L k \frac{b^2-1}{2b} + \frac{m}{4} L^2$ for odd $b$.
For odd $m$ the last item can be substituted with $\frac{m^2-1}{4m} L^2$ so that for $m=b$ and odd $b$ the total length of entire paths does not exceed $(2(L-1)Lk + L^2)\frac{b^2-1}{4b}$. Hence the upper bound for the average distance reads
$$ 
  d \leq \frac{bk}{2} + \frac{m}{4}\frac{L}{L-1} \mbox{for even} b,
$$
$$
  d \leq \frac{(b^2-1)k}{2b} + \frac{m}{4}\frac{L}{L-1} \mbox{for odd} b.
$$
In particular, for $m=b$ and even $b$: 
$$
d \leq \frac{b}{4} \left( 2 \log_b L - 1 + \frac{1}{L-1} \right) \ ,
$$
for $m=b$ and  odd $b$: 
$$
d \leq \frac{b^2-1}{4b} \left( 2 \log_b L - 1 + \frac{1}{L-1} \right) \ .
$$

\paragraph{LNS constructed by the algorithm D4.} Similarly to the case D4s for the average distance can be found upper bound by consideration of paths for all pairs of nodes. In the case $L=mb^k$, $k\leq b$ the estimations reads
\begin{eqnarray}
d &\leq & \frac{bk}{2} + \frac{m}{4}\frac{L}{L-1} +2k-2 \qquad\mbox{ for even } b\ ,\nonumber\\
\label{SubCircEstim}\\
d &\leq & \frac{(b^2+1)k}{2b} + \frac{m}{4}\frac{L}{L-1} +2k-2 \qquad\mbox{ for odd } b \nonumber
\end{eqnarray}
(the term $2k-2$ is added because for transition from the cycle of the level $i$ to the cycle of the level $i+1$ for $i=1,\ldots,k-1$ one has to make an additional step along an edge of the underlying lattice).

If the conditions $L=mb^k$ and $k \leq b$ is not fulfilled, lengths of first-level shortcuts and the number of steps over shortcuts of a level $i$ necessary for reaching the node $i+v_{i+1} b^{i+1}$ which is next to the node $i+1+v_{i+1} b^{i+1}$ entering the $i+1$-level cycle, may prove to be greater then $b$. However these differences affect a small part of the nodes. We suppose, on the analogy with (\ref{SubCircEstim}) that in this case the following upper bound is correct:
\begin{equation}
d \lesssim \lambda\left[\frac{k(b+4)}{2} + \frac{L}{4b^k} - 2\right]\ ,             \label{D4_estim}
\end{equation}
where $\lambda\simeq1$.

\begin{figure}
\begin{center}
\includegraphics[scale=.4]{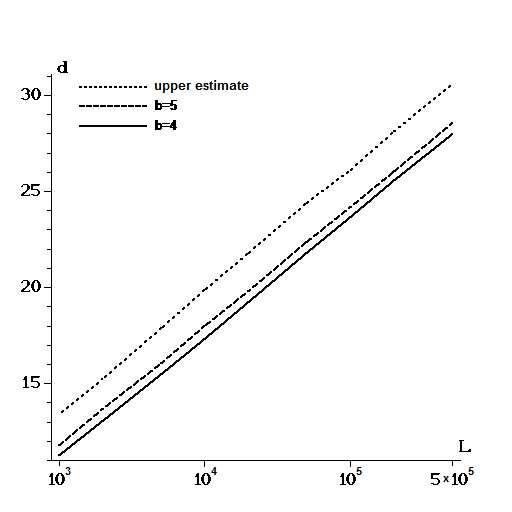}
\end{center}
\caption{Average distances $d$ in D4-networks with $b=4,\ 5$ for $L=1000,\ 2000,\ 5000,\ 10000,\cdots,\ 500000$; $k$ is the maximum integer such that $b^k<L/2$; the dotted line is the upper bound (\ref{D4_estim}) at $\lambda=1$}
\label{fig:d_for_D4_b_4_5_and_estim}
\end{figure}

From this estimation it follows that with an appropriate choice of the parameters the average distance in the subcirculant networks grows no faster than $\log\,L$. In particular, $d\lesssim 4(u-2) + 2 = 2 \log_2 L - 6$ for $L=2^{2u}$, $b=4$, $k=u-2$. Numerical simulations confirm this conclusion: the results for $d$ in D4-networks with 
$b=4,\ 5$ for $L=1000,\ 2000,\ 5000,\ 10000,\cdots,\ 500000$ is depicted in fig.~\ref{fig:d_for_D4_b_4_5_and_estim}, the parameter $k$ being maximal integer such that $b^k<L/2$. The estimation of the upper bound (\ref{D4_estim}) for $\lambda=1$ is also presented in this figure. It is seen that the average distance indeed logarithmically grows with growth of the network size and the estimation obtained correctly reflects this dependence. 

As it was mentioned in section \ref{DetAlg}, multiplicative circulant graphs also possess the small-world properties (cf. (\ref{circDiam})). A comparison of the circulant graphs and suggested in our work subcirculant graphs will be presented in section \ref{OAKS}.

\section{Local navigation in LNSs\label{LNRP}}

As it was noticed in Introduction, from the point of view of optimization of parameters of stochastic LNSs the dependence of navigation length on parameters of the algorithms is of special interest. Pioneering work in the field of local navigation in spatially embedded networks with small-world properties is \cite{K2000} by Kleynberg while the very formulation of the problem goes back to the classical Milgram's experiment\cite{Mil} .

In short the navigation problem in LNSs is set as follows:
\begin{itemize}
\item a node ``knows'' only position of the destination node in the underlying lattice and all its neighbors with the account of shortcuts (in more elaborated versions the node may ``know'' its neighbors with the depth more than unity);
\item the problem is to find the shortest way to destination using only this local information (without knowing all the global LNS structure, in other words without knowledge about positions of all the shortcuts).
\end{itemize}
The simplest local navigation follows the greedy algorithm: an entity goes from a node to that of its neighbors which geographically (i.e., in the sense of coordinates on the underlying lattice) is closest to the destination node.

The most important characteristic property of the navigation process is the average navigation path length between nodes of a network which is defined similarly to (\ref{SDEA03}), (\ref{SDDA01}):
\begin{equation}
\ell= \frac{2}{L(L-1)}\sum_{i>j}\left\langle {\ell}_{ij}\right\rangle=\frac{1}{L(L-1)}\sum_{i,j}\left\langle {\ell}_{ij}\right\rangle\ ,                                                                       \label{LNRP01} 
\end{equation}
where $\ell_{ij}$ the navigation length between nodes $i$ and $j$. Of course for deterministic algorithms the averaging over ensemble $\langle\cdot\rangle$ is absent.

\subsection{Navigation path length for stochastic algorithms\label{NDSA}}
\begin{figure}
\begin{center}
\includegraphics[scale=0.5]{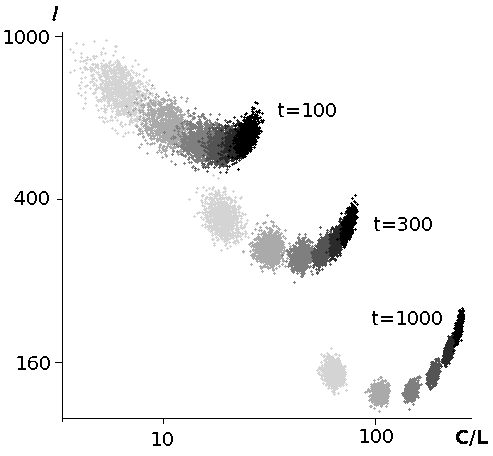}
\end{center}
\caption{Dependence of the distribution of the average (over instances) navigation path length $\ell$ on shortcut length at three different values of  $t$ and for $\alpha=0,\ 0.2,\ 0.4,\ 0.6,\ 0.8,\ 1.0$; the lighter the shading the larger value of $\alpha$}
\label{fig:ZavNavDliny_ot_ceny}
\end{figure}

The dependence of distributions of the averaged (over instances) navigation path lengths $\ell$ on shortcuts length per node as well as on the parameters $t$ and $\alpha$ in the case of the algorithm S1m is shown in fig.~\ref{fig:ZavNavDliny_ot_ceny}. For S1 the behaviour is similar when $t$ is substituted by $p$. It is seen that there exist a minimum of $\ell$ at different values of $\alpha$ depending on $t$. Predictably the more shortcuts the smaller average navigation path length in the minimum. It is seen also that for a given cost one can find the parameters of the algorithm S1m with smallest navigation path length and vice versa for given navigation path length one can find the parameters with smallest total shortcuts length per node. Therefore in this case one has to solve the cost-quality optimization problem too. 

\begin{figure}
\begin{center}
\includegraphics[scale=.25]{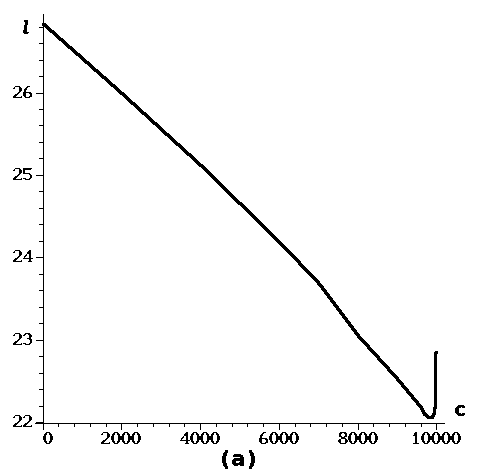}
\hspace{5mm}
\includegraphics[scale=.25]{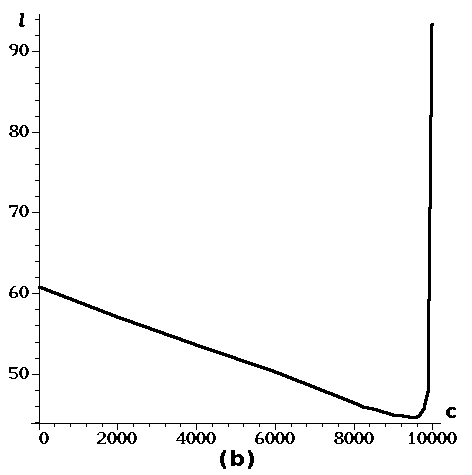}
\hspace{5mm}
\includegraphics[scale=.25]{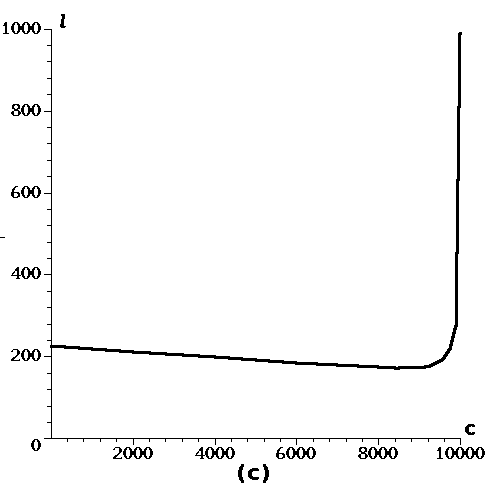}
\end{center}
\caption{Dependence of the average navigation path lengths on $c$ for the algorithm S2;  $t=L=10^4$; (a): $\alpha=1$ , (b): $\alpha=1.5$, (c): $\alpha=2$}
\label{fig:NavDlinAlgS2}
\end{figure}

Decrease of the average navigation path length for S2 with growth of the number $c$ of forcibly conjoined shortcuts is presented in fig.~\ref{fig:NavDlinAlgS2}. As it will be shown in section \ref{OAKS},  S2 has the best cost-quality ratio among all the stochastic algorithms.

\subsection{Navigation path length for hierarchical deterministic algorithms\label{NDRI}}
\begin{figure}
\begin{center}
\includegraphics[scale=.4]{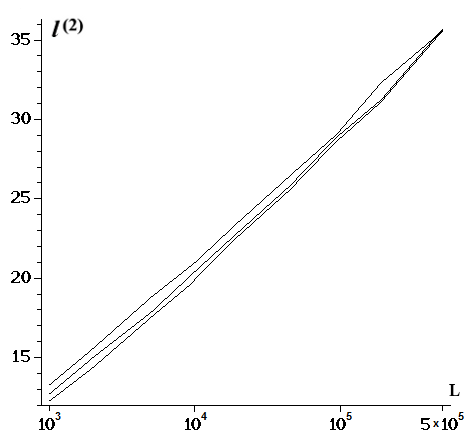}
\end{center}
\caption{Dependence of the average path length $\ell^{(2)}$ for the two-level navigation on logarithm of the number of nodes; $b=4$ (lower curve), $5$, (middle curve) $6$ (upper curve); $k$ is chosen to minimize $\ell^{(2)}$ for a given $L$}
\label{fig:d2_for_SubC}
\end{figure}

Numerical simulations show that the global average distance between nodes in the networks constructed by algorithms D2, D4s and D4 is very small. However such networks possess rather large average path length in the case of the local navigation based on the elementary greedy algorithm. Accordingly, in this work the modified algorithm of local navigation, namely the two-level local navigation, is suggested. In this modification not only the nearest neighbors of a given node are taken into account but also  neighbors of neighbors. Thus in the case of an interconnection network a message from a given node is transferred to that its neighbor which in turn has a neighbor closest to the destination node in the sense of the underlying lattice. The two-level greedy algorithm requires a little bit more calculations at each node but still remains local and well scalable. The average path length in case of two-level local navigation is denoted by $\ell^{(2)}$. Notice that the global average path length $d$ can be considered as a limiting case of navigation length with the infinite depth of search. The results of numerical simulations for $\ell^{(2)}$ in LNSs constructed by the algorithm D4 is shown in fig.~\ref{fig:d2_for_SubC}.

\section{Selecting the algorithm for LNS construction\label{OAKS}}

\subsection{Multicriteria cost-quality optimization\label{MOCK}}
Multicriteria (or multiobjective) optimization has been applied in many fields of science where optimal decisions need to be taken in the presence of trade-offs between two or more conflicting objectives. In our case we will consider two criteria: average path length between nodes (global $d$ or navigation $\ell,\ \ell^{(2)}$) that expresses ``quality'' of an LNS, and shortcuts length per node $C/L$ that reflects ``cost'' of creation of the LNS. From the consideration in the preceding sections it is obvious that these two quantities are interrelated: by increasing the cost $C/L$ it is possible to improve quality (to reduce $d,\ \ell$ or $\ell^{(2)}$), and, conversely, improving the quality causes increasing the cost.

There exist a number of approaches to multicriteria optimization (see, e.g. \cite{Ste}). We will use the method of weighted sum (in more general context, scalarizing multiobjective optimization problems). For this aim we define the following target functions: 
\begin{eqnarray}
G_w&=&wd+(1-w)C/L\ ,          \label{OAKS01}\\[3mm]    
G^\prime_w&=&w\ell+(1-w)C/L\ ,          \label{OAKS02}\\[3mm]
G^{\prime\prime}_w&=&w\ell^{(2)}+(1-w)C/L\ .          \label{OAKS02a}    
\end{eqnarray}

Minimization of these target functions means that optimal values for parameters of the algorithms are picked out from point of view of quality (small average path length) and cost (total shortcuts length per node). With such a minimization the parameter $0\leq w\leq1$ characterizes the relative significance of the each of the criteria (cost and quality). In other words, the suggested method of optimization implies that for each value of the significance parameter $w$ one have to find values of the algorithm parameters (e.g., $p/t$ and $\alpha$ for S1/S1m;  $t,\ c$ and $\alpha$ for S2; $b$ and $k$ for D4; the network size $L$ is supposed to be fixed) which minimizes  $G_w$, $G^\prime_w$ and $G^{\prime\prime}_w$.

\begin{figure}
\begin{center}
\includegraphics[scale=.4]{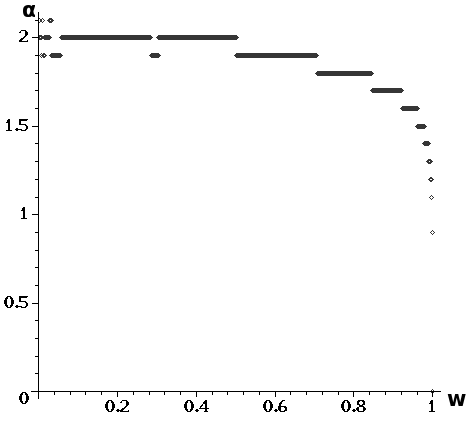}
\includegraphics[scale=.4]{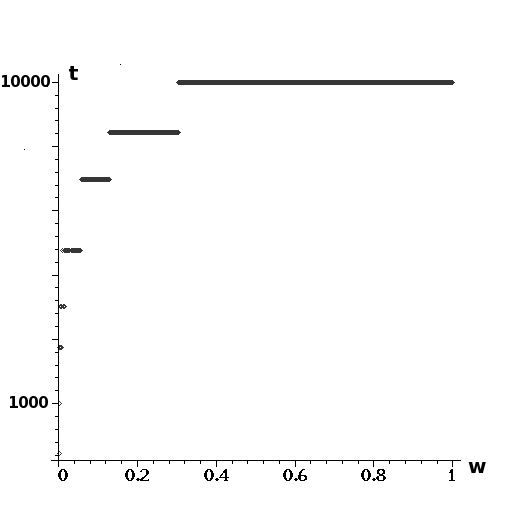}
\end{center}
\caption{Dependence of optimal values of $\alpha$ (a) and $t$ (b), minimizing $G_w$, on the quality significance parameter $w$ ($L=10^4$)}
\label{fig:G_w_minimizing_S1m}
\end{figure}

To illustrate an application of this method for LNS analysis the values of the parameters $t$ and $\alpha$ which minimize $G_w$ are presented in fig.\ref{fig:G_w_minimizing_S1m}  for network ensembles constructed by S1m. These results imply that when the quality significance $w$ is greater than ~ 0.35, the number of the shortcuts must be maximum possible (equal to the number of nodes) while at smaller values of $w$ the importance of shortcuts length minimization leads to decreasing the number of shortcuts (values of $t$). On the contrary, values of $\alpha$ close to 2 (the critical value for one-dimensional small-world LNS \cite{MM}) are optimal ones for $w$ smaller than ~ 0.5, that is in the domain of prevailing cost.

\begin{figure}
\begin{center}
\includegraphics[scale=.4]{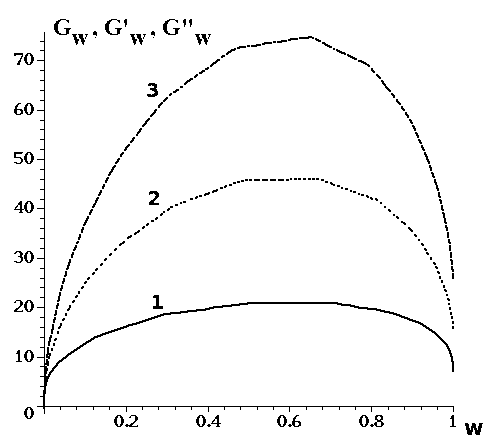}
\end{center}
\caption{Dependence of minimum values of the target functions on $w$ in the case of algorithm S1m: the curve 1~--- min $G_w$; the curve 2~--- min  $G^{\prime\prime}_w$ ; the curve 3~--- min $G^\prime_w$}
\label{fig:G_w_G1_w_G2_w_for_S1m}
\end{figure}

Results of calculation of minimum values of $G_w$, $G^\prime_w$ and $G^{\prime\prime}_w$ for the S1m LNSs are presented in fig.~\ref{fig:G_w_G1_w_G2_w_for_S1m} ($L=10^4$; averaging over ensembles of 100–1000 instances). Each point on the curves corresponds to an ensemble with parameters $t,\ \alpha$ providing minimum $G_w$, $G^\prime_w$ and $G^{\prime\prime}_w$ for a given $w$. It is seen that in the total range of $w$ the function $G_w$ (``infinite'' searching depth) has smallest values, the values of  $G^{\prime\prime}_w$ (searching depth is equal to 2) is greater than those for $G_w$ and the values of $G^\prime_w$ (searching depth is equal to 1) are the largest ones. This result is intuitively understandable: the more information one uses for searching a shortest path the smaller values of the target function.

More important that the target functions $G_w$, $G^\prime_w$ and $G^{\prime\prime}_w$ make possible to compare different algorithms of LNS construction including those with different set of parameters.

\subsection{Simulation results for different algorithms\label{RCOA}}
\begin{figure}
\begin{center}
\includegraphics[scale=.4]{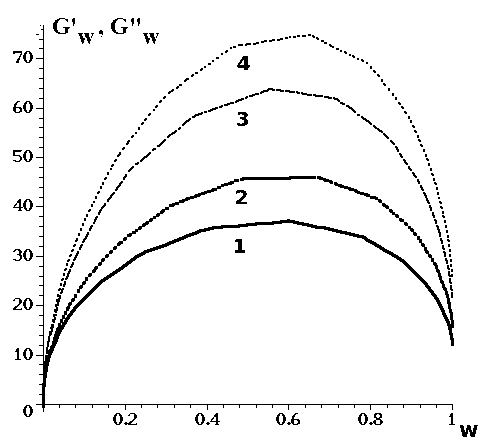}
\end{center}
\caption{Dependence of minimum values of the target functions on $w$: the curve 1~--- min $G^{\prime\prime}_w$ for S2; 2~--- min $G^{\prime\prime}_w$ for S1m; 3~--- min $G^\prime_w$ for S2; 4~--- min $G^\prime_w$ for S1m}
\label{fig:G1_w_and_G2_w_for_S1m_and_S2}
\end{figure}

Results of the comparison of two stochastic algorithms is presented in fig.~\ref{fig:G1_w_and_G2_w_for_S1m_and_S2}. They show that with the use both of two-level and ordinary (one-level) navigation, S2 is the preferred algorithm.

\begin{figure}
\begin{center}
\includegraphics[scale=.4]{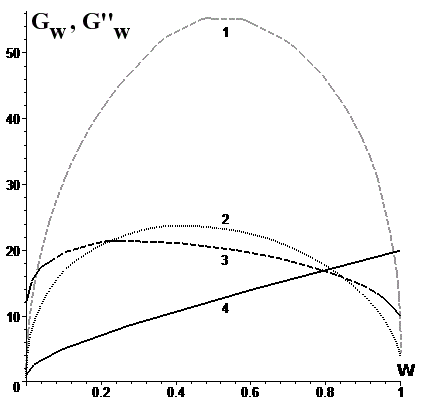}
\end{center}
\caption{The dashed curve 1~--- min $G_w$ for the multiplicative circulant graphs; the dotted curve 2~--- min $G_w$ for D3 with star-like connection of hubs; the dashed curve 3~--- min $G_w$ for D3 with double-loop connection of hubs; solid curve 4~--- min $G^{\prime\prime}_w$ for D4; $L=10^4$}
\label{fig:G_w_for_D3_and_circulant_vs_G2_w_for_D4}
\end{figure}

A comparison of minimal values of the target functions for the algorithms D3, D4 as well as for the multiplicative circulant graphs is shown in fig.~\ref{fig:G_w_for_D3_and_circulant_vs_G2_w_for_D4}. It is seen that if the cost is of the essential significance ($w\lesssim 0.8$), D4 has advantage over D3 and the circulant graphs. On the contrary in the case of prevailing quality ($w\gtrsim 0.8$), D3 may be more effective especially with star-like connection of hubs. This is not surprising because the star-like graph has quite small diameter ${\cal D}=2$.However we have mentioned already that the star-like configuration is too fragile since in case of a failure of the central node the structure proves to be completely disintegrated.

\begin{figure}
\begin{center}
\includegraphics[scale=.4]{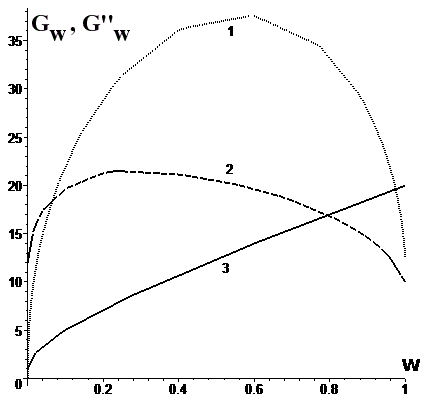}
\end{center}
\caption{Dependence of minimum values of the target functions on $w$: the dotted curve 1~--- min~$G^{\prime\prime}_w$ for S2; the dashed curve 2~--- min~$G_w$ for D3 with double-loop connection of hubs; the solid curve 3~--- min~$G^{\prime\prime}_w$ for D4; $L=10^4$}
\label{fig:Gw_G1w_S2_D4}
\end{figure}

Thus D4 is the preferred algorithm in the class of deterministic algorithms in wide range of the parameter of quality significance $w$. A comparison with stochastic algorithms remains to carry out. An appropriate data are presented in fig.~\ref{fig:Gw_G1w_S2_D4}, where minimal values of the target functions for S2, D3 (with double loop connection of the hubs), and D4 are shown. The results show that practically in the whole range of the parameter $w$ the deterministic algorithms D3 and D4 are preferred.

\begin{figure}
\begin{center}
\includegraphics[scale=.4]{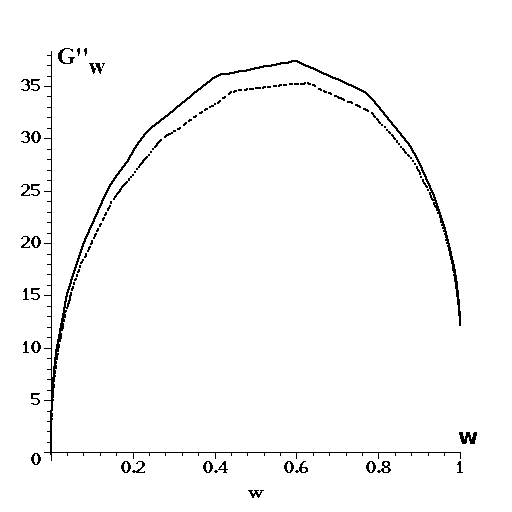}
\end{center}
\caption{Dependence of minimum values of the target functions $G^{\prime\prime}_w$ on $w$ for the algorithm S2: solid curve~---for characteristics averaged over ensemble; dashed curve~--- for instances from the ensemble having in a sample of 100 instances the fifth in the order of increasing averaged path length between nodes (fifth percentile)}
\label{fig:G2_w_for_S2_average_and_5p}
\end{figure}

It is worth noticing that the comparison with the stochastic algorithms was carried out for the characteristics averaged over ensembles of LNSs. One may wonder if selecting instances with good characteristics may result in essential improvement of the cost-quality ratio. The numerical simulation shows that in the case of the best stochastic algorithm S2 such an improvement is hardly possible, see fig.~\ref{fig:G2_w_for_S2_average_and_5p}.

\section{Conclusion\label{Zak}}

In this work we considered a number of both known in the literature and original algorithms for construction of complex networks with the property of a small world, namely with slow (logarithmic) growth of average distance between nodes with growth of the network size. The networks constructed on the basis of these algorithms have basic structure of a regular lattice with additional shortcuts providing the small-world properties. We suggested a method for comparison of various algorithms on the basis of optimization of the cost-quality ratio. Under the character of ``cost'' we have chosen unit wiring cost, i.e., total length of all shortcuts divided by the number of nodes in the network (total shortcuts length per node), while ``quality'' is the global or navigation average path length between nodes. The optimization is carried out for a weighted sum in which the relative significance of the cost and quality is defined by a value of an appropriate parameter (weight).

The two main classes of the algorithms considered are stochastic and deterministic ones. Therefore, an important question is which type of the algorithms has better cost-quality ratio. As a result of the investigations it has been shown that in the case of one-dimensional underlying lattice and from the point of view of the cost-quality ratio the preferred networks are proved to be the subcirculant ones which are constructed by means of the deterministic algorithm (denoted in this work as D4). The subcirculant networks have the structure similar to that of the well known multiplicative circulant graphs but with essentially smaller number of shortcuts (and, respectively, lower wiring cost).

It necessary to stress that though in this work the principal use case was designing an optimal interconnection networks for future generation supercomputers, we believe that the results obtained can find applications in many other areas when communication properties of some regular network need to be improved without destroying the underlying basic structure. 

In the subsequent publications we will continue the consideration of the lattice networks with shortcuts (LNS) both from the point of view of other characteristics (in particular, beweeness centrality, fault tolerance, etc.) and generalizations to higher dimensional underlying lattices.

\vspace{5mm}

This work is partially supported by RFBR grant No. 12-07-00408-a.


\end{document}